\documentclass{article}
\usepackage{amsmath,amssymb}
\newtheorem{definition}{Definition}

\newtheorem{lemma}{Lemma}
\newenvironment{proof}{%
\par\noindent\textbf{Sketch of proof.}\hspace{1ex}}%
{\hfill$\Box$\par\medskip}
\usepackage{fullpage}
\advance\hoffset by -3mm  
\advance\voffset by  8mm  
\def\invariant{\mathfrak}
\newcounter{rem}
\newenvironment{remark}{%
  \refstepcounter{rem}%
  \par\medskip\noindent\textbf{Remark \therem\: ---}\hspace{1ex}}%
{\par\medskip}
\newcounter{prop}
\newenvironment{proposition}{%
  \refstepcounter{prop}%
  \par\medskip\noindent\textbf{Proposition \theprop\: ---}\hspace{1ex}}%
{\par\medskip}
\newcounter{thm}
\newenvironment{theorem}{%
  \refstepcounter{thm}%
  \par\medskip\noindent\textbf{Theorem \thethm\: ---}\hspace{1ex}}%
{\par\medskip}
\newcounter{ex}
\newenvironment{example}{%
  \refstepcounter{ex}%
  \par\medskip\noindent\textbf{Example \theex\: ---}\hspace{1ex}}%
{\par\medskip}
\title{Polynomial Time Nondimensionalisation \\of Ordinary Differential
  Equations \\via their Lie Point Symmetries
}

\author{%
  \begin{tabular}{cc}
  \'Evelyne Hubert &  Alexandre Sedoglavic  \\
  {Projet CAF\'E, INRIA Sophia Antipolis} & {Projet ALIEN,  INRIA Futurs \&} \\
  {BP~93, F-06902 Sophia-Antipolis, France}  & {LIFL UMR 8022 CNRS, USTL}\\
  {Evelyne.Hubert@inria.fr} &   {F-59655 Villeneuve d'Ascq, cedex, France}\\
&{Alexandre.Sedoglavic@lifl.fr}  
  \end{tabular}
}
\date{February~2006}

\begin{document}
\maketitle
\begin{abstract}
  Lie group theory states that knowledge of a~$m$-parameters solvable
  group of symmetries of a system of ordinary differential equations
  allows to reduce by~$m$ the number of equation.
  \par
  We apply this principle by finding dilatations and translations that
  are Lie point symmetries of considered ordinary differential system.  By
  rewriting original problem in an invariant coordinates set for these
  symmetries, one can reduce the involved number of parameters. This
  process is classically call nondimensionalisation.
  \par
  We present an algorithm based on this standpoint and show that its
  arithmetic complexity is polynomial in input's size.  
\end{abstract}
\section{Introduction}
\label{sec:Introduction}
This paper is devoted to the process of nondimensionalisation that it
described in~\cite{Murray2002} as follow: \textit{``Before analysing
  [a] model it is essential, or rather obligatory to express it in
  \emph{nondimensional} terms. This has several advantages. For
  example, the units used in the analysis are then unimportant and the
  adjectives small and large have a definite relative meaning. It also
  always reduces the number of relevant parameters to
  \emph{dimensionless groupings} which determine the dynamics''.}
\begin{example}  \label{ex:VerhulstLinearPredation}
  In order to illustrate this statement, let us consider the following
  Verhulst's logistic growth model with linear predation (see~\S~1.1
  in~\cite{Murray2002}):
  \begin{equation}
    \label{eq:VerhulstLinearPredation}
    {\textup{d}x}/{\textup{d}t} = x(a-bx) - c x, \quad \dot{a}=\dot{b}=\dot{c}=0,
  \end{equation}
  for which all forthcoming computation could be easily performed by
  hand.
  \par\noindent\textbf{Step~1.}  
  One can remark that the following one parameter group of translation
  symmetries:
  \begin{equation}
    \label{eq:TranslationVerhulstLinearPredation}
\mathcal{T}_{\lambda}:    a \rightarrow a + \lambda, \quad c \rightarrow c + \lambda
  \end{equation}
  and the following two parameters group of scale symmetries:
  \begin{equation}
    \label{eq:ScalingVerhulstLinearPredation}
    \mathcal{S}_{(\mu,\nu)}:
    \begin{array}{ccr} 
      t & \rightarrow & t/\nu, \\
      x & \rightarrow & \mu x,
    \end{array} 
    \qquad
    \begin{array}{ccc} 
      a & \rightarrow & \nu a, \\
      b & \rightarrow & \nu b/\mu, \\
      c & \rightarrow & \nu c,
    \end{array} 
  \end{equation}
  leave invariant solutions of
  system~(\ref{eq:VerhulstLinearPredation}).
  \par\noindent\textbf{Step~2.}  
  Assuming that~${a\not =c}$, one can determine some rational
  invariants of these groups like:
  \begin{equation}
    \label{eq:InvariantsVerhulstLinearPredation}
    \invariant{t} = (a-c)\,t,\qquad
    \invariant{x} = \frac{b}{a-c}\,x.
  \end{equation}
  These invariants are the \emph{dimensionless groupings} evoked in
  above citation.
  \par\noindent\textbf{Step~3.}  
  Using above rational invariants,
  system~(\ref{eq:VerhulstLinearPredation}) can be rewritten in these
  new coordinates as follow:
  \begin{equation}
    \label{eq:ReducedVerhulstLinearPredation}
    \textup{d}\invariant{x}/{\textup{d}\invariant{t}} = \invariant{x}(1-\invariant{x}). 
  \end{equation}
  In fact, one can deduce from~(\ref{eq:VerhulstLinearPredation})
  and~(\ref{eq:InvariantsVerhulstLinearPredation}) a differential
  system that could be simplified in order to
  obtain~(\ref{eq:ReducedVerhulstLinearPredation}).  Thus,
  groups~(\ref{eq:TranslationVerhulstLinearPredation})
  and~(\ref{eq:ScalingVerhulstLinearPredation}) allow to rewrite our
  original problem with a reduced number of parameters.
\end{example}
We have considered a variety of such dynamical systems that arise from
biological, physical, etc.\ models. As often observed, those systems
can be rewritten in a lower number of variables or parameters than
initially appears.  
\par
There is no difficulty to perform manipulations done above by hand on
simple systems. But while system's complexity increases, these
manipulations require more avoidable work as illustrated by the
following example.
\begin{example}  \label{ex:Murray2002p88}
  Let us consider the following prey-predator model taken
  from~page~$88$ in~\cite{Murray2002}:
  \begin{equation}
    \label{eq:Murray2002p88}
    \left\lbrace
      \begin{array}{cll} 
        \frac{\textup{d}n}{\textup{d}t} & = & 
        \left(\left(1-\frac{n}{k_{1}}\right)r-k_{2}\frac{p}{n+e}\right)n,
        \\[\medskipamount]
        \frac{\textup{d}p}{\textup{d}t} & = & 
        \left(1-h\frac{p}{n}\right) p\, s,
        \\[\medskipamount]
        \multicolumn{3}{l}{\dot{r}=\dot{s}=\dot{e}=
          \dot{h}=\dot{k}_{1}=\dot{k}_{2}=0.}
      \end{array} \right.
  \end{equation}
  \textbf{Step~1.}  One can determine that the following~$3$
  parameters group of scale symmetries:
  \begin{equation}
    \label{eq:ScalingMurray2002p88}
    \!
    \mathcal{S}_{(\lambda,\mu,\nu)}: \begin{array}{ccr} 
      t & \rightarrow & t/\lambda, \\
      n & \rightarrow & \mu n,\\
      p & \rightarrow & \mu p/\nu,
    \end{array} 
    \begin{array}{ccc} 
      r & \rightarrow & \lambda r,\\
      s & \rightarrow & \lambda s, \\
      e & \rightarrow & \mu e,
    \end{array} 
    \begin{array}{ccc} 
      h & \rightarrow & \nu h,\\
      k_{1} & \rightarrow & \mu k_{1},\\
      k_{2} & \rightarrow & \lambda\nu k_{2}
    \end{array} 
  \end{equation}
  leaves invariant solutions of system~(\ref{eq:Murray2002p88}).
  Computations show that this system does not have any translation
  symmetries.
  \par\noindent
  \textbf{Step~2 \& 3.}  If group's parameters are specialized as
  follow:
  \begin{equation}
    \label{eq:SpecializationMurray2002p88}
    \lambda= \frac{1}{r},\quad \mu = \frac{1}{k_{1}},\quad \nu = \frac{r}{k_2},
  \end{equation}
  resulting transformation~$\mathcal{S}_{(1/r,1/k_{1},r/k_{2})}$
  induces a change of coordinates that send coordinates~$r,\ k_{1}$
  and~$k_{2}$ to~$1$ and other coordinates to rational invariants:
  \begin{equation}
    \label{eq:InvariantsMurray2002p88}
    \invariant{t} = r\,t,\ \
    \invariant{n} = \frac{n}{k_{1}},\ \
    \invariant{p} = \frac{k_{2}p}{k_{1}r},\ \
    \invariant{s} = \frac{s}{r},\ \
    \invariant{e} = \frac{e}{k_{1}},\ \
    \invariant{h} = \frac{r h}{k_{2}}.
  \end{equation}
  After applying transformation~$\mathcal{S}_{(1/r,1/k_{1},r/k_{2})}$
  on system~(\ref{eq:Murray2002p88}), we obtain a new system expressed
  in above new coordinates:
  \begin{equation}
    \label{eq:ReducedMurray2002p88}
    \left\lbrace
      \begin{array}{cll} 
        \frac{\textup{d} \invariant{n}}{\textup{d}\invariant{t}} & = & 
        \left(1-\invariant{n}-\frac{\invariant{p}}{\invariant{n} + 
          \invariant{e}}\right)\invariant{n},
        \\[\medskipamount]
        \frac{\textup{d}\invariant{p}}{\textup{d}\invariant{t}} & = & 
        \left(1-\invariant{h}\frac{\invariant{p}}{\invariant{n}}\right)
        \invariant{p}\, \invariant{s},
        \\[\medskipamount]
        \multicolumn{3}{l}{\dot{\invariant{s}}=\dot{\invariant{e}}
          =\dot{\invariant{h}}=0.}
      \end{array} \right.
  \end{equation}
  and as in example~\ref{ex:VerhulstLinearPredation}, we reduce by~$3$
  the number or parameters. Thus, any further manipulations of
  system~(\ref{eq:VerhulstLinearPredation}) (for example, its phase
  plane analysis or study of its bifurcation analysis) is simplified in
  the new set of coordinates~(\ref{eq:InvariantsMurray2002p88}).
  \par
  Same process could be achieved in another set of invariants
  coordinates by choosing to send~$s,\ e$ and~$h$ to~$1$ using
  transformation~$\mathcal{S}_{(1/s,1/e,1/h)}$.
\end{example}
\begin{remark} 
  Example~\ref{ex:Murray2002p88} shows how scale symmetries allow
  computation of a set of invariant coordinates and of resulting new
  system by an evaluation without any further algebraic manipulation
  (see Section~\ref{sec:ComputationOfRationalInvariants}).
\end{remark}
\subsection{Dimensional Analysis and some Lie point Symmetries}
\label{sec:DimensionalAnalysis}
There is just seven primary units in the International Metric System
and all physically meaningful equations could be written in these
units. Such physical \emph{primary dimension} (say time) can be
expressed in different units (second, hour, etc.) and thus, these
units can be scaled independently of each other~(a hour is~$3600$
seconds, etc).  This possibility induces also dilatations on
\emph{secondary dimension} like speed ($m/s$) or corporal mass index
($kg/m^2$), etc.  Thus, there is likely some scale transformations
group acting on functional relation among these kind of quantities.
\par
Dimensional analysis is based on this remark.  Bridgman explains
in~\cite{Bridgman1922}: \textit{``The principal use of dimensional
  analysis is to deduce from a study of the dimensions of the
  variables in any physical system certain limitations on the form of
  any possible relationship between those variables''.}  Thus,
dimensions analysis addresses very general problems and we consider in
this paper just one of its consequences that is nondimensionalisation.
This process is done because reducing the number of parameters, or
possibly also the number of state variables, is an advantage for
studying qualitative features of the model.
\paragraph{Previous related works.}
For large systems, nondimensionalisation could become a difficult
process that motivate several implementations (see~\cite{Khanin2001}
and the references therein for more details). Up to our knowledge,
there is no complexity result concerning these works that are related
to~$\Pi$ theorem and to rules of thumbs based on the knowledge of
units in which is expressed the problem.
\par
As we notice at beginning of this section, there is often some scale
symmetries group of ordinary differential system describing
biological, physical, etc.\ phenomena. This fact could be considered
in the framework of Lie symmetries group theory (see~\S~3.4
in~\cite{Olver1993} for such presentation of~$\Pi$ theorem). This
standpoint allows to presents computation involved in
nondimensionalisation in a very simple and efficient way that we did
not found in literature.
\par
The present study was motivated by this fact and by authors' inability
to apply classical dimensional analysis' rules to systems composed of
more then~$7$ equations and a dozen parameters.
\begin{remark} 
  In this paper, we restrict ourself to translation and scale
  symmetries that occurs frequently in application. Same type of
  result could be obtain for other kind of Lie point symmetries (rotation,
  inversion, etc. See Example~\ref{ex:FitzHughNagumo} in
  Section~\ref{sec:Conclusion}).
\end{remark}
\subsection{Main contribution}
\label{sec:MainContribution}
It is the aim of this paper to provide an algorithm to make the
reduction evoked in previous section.  In the example above
and~$90$\% that we tried, the system~(\ref{eq:ODS}) is actually
symmetric under a group of scalings and/or translations.  New
variables that we introduced in above examples and many others
actually form a generating set of rational invariants for some group
action.
\par
Computing symmetry of a differential system has been an intensive
field of application of computer algebra, especially to mathematical
physics~\cite{Anderson1996, ChebTerrabRoche1998, Olver1993} and
computing a generating set of (differential) invariants for general
group action has received relatively recently firm
foundations~\cite{HubertKogan2006, Olver1999, Olver1993}.
\par
The viewpoint of this paper is not to be general nor theoretical but
practical for a large class of problems. We shall apply known general
theory to special cases. We obtain efficient algorithms for reducing
the number of parameters in biological, chemical, etc.\ models basing
ourselves on the observation of a general scenario: the invariance of
the model under a group of scaling and translation.  We provide
efficient computer algebra algorithms for computing the scaling and
translation symmetry of a differential system~(\ref{eq:ODS}), compute
their invariants and rewrite the system in terms of those.  We thus
obtain the reduced system. These result are summarized in the
following statement:
\begin{theorem} 
  \label{thm:MainResult}
  Let~$\Sigma$ be a differential system bearing on~$n$ state variables
  and depending on~$\ell$ parameters that is coded by a straight-line
  program of size~$L$ (see Section~\ref{sec:MathematicalFramework}).
  There exists a probabilistic algorithm that determines if a Lie
  point symmetries group of~$\Sigma$ composed of dilatation and translation
  exists; in that case, a rational set of invariant coordinates is
  computed and~$\Sigma$ is rewrite in this set with a reduced number
  of parameters.
  \par 
  The arithmetic complexity of this algorithm is bounded by
  $$
  \mathcal{O}\Big((n+\ell+1)\big(L+(n + \ell + 1)(2n +\ell+1)\big)\Big).
  $$
\end{theorem}
\paragraph{Outline of the paper.}  In the next
section, we recall some basic definitions of differential algebra and
we present in this framework the relationship between
dilatation/translation transformations and induced derivations that
are theirs infinitesimal generators.  Then, we recall that such
infinitesimal generators are defined by a partial differential
equations system i.e.\ some infinitesimal conditions which are
presented and used in the sequel.  
\par
In the last part of this paper, we show that infinitesimal conditions
allow by a Gaussian elimination performed on a constant field to
determine Lie groups of scale/translation symmetries.  In fact, for
these transformations, infinitesimal conditions split into a linear
system of algebraic equations.  We point out that computation of sets
of rational invariant coordinates and rewriting of original system in
these sets is reduced to linear algebra and we estimate the related
arithmetic complexity.
\section{Mathematical framework}
\label{sec:MathematicalFramework}
Hereafter, we consider an algebraic ordinary differential system
bearing on~$n$ state variables~${X:=(x_{1},\dots,x_{n})}$ and
depending on~$\ell$
parameters~${\Theta:=(\theta_{1},\dots,\theta_{\ell})}$:
\begin{equation}
\label{eq:ODS} 
\Sigma \qquad \left\lbrace
\begin{array}{l} 
\dot{t}  =  1, \quad \dot{\Theta}  =  0, \\ 
\dot{X}  =  F(t,X,\Theta). 
\end{array} \right.
\end{equation}
The letter $\dot{X}$ stands for first order derivatives of state
variables~${(\dot{x}_{1}, \dots, \dot{x}_{n})}$ w.r.t.\ time~$t$; we
use the standard notation~${X^{(i)}=\big({{x}_{1}}^{\!(i)}, \dots,
  {{x}_{n}}^{\!(i)}\big)}$ for higher derivatives of order~$i$.  We
assume that~${F:=(f_{1},\dots,f_{n})}$ consist of rational functions
over a subfield~$\mathbb{K}$ ($\mathbb{Q}$ for example)
of~$\mathbb{C}$ i.e.~$F$ is a finite subset
of~$\mathbb{K}(t,X,\Theta)$.
\paragraph{Complexity model.}
We evaluate the complexity of our algorithms within the model of
\emph{straight-line program} (see~{\S}~4
in~\cite{BurgisserClausenShokrollahi1997}).  For instance the
expression~${e := (x+1)^{5}}$ is represented by the following kind of
instructions sequence:
\begin{equation}
  \label{eq:ExampleSLP}
  {e_{1}:= x+1,\quad e_{2}:= {e_{1}}^{\!2},\quad e_{3}:={e_{2}}^{\!2},\quad
    e:=e_{3} e_{1}}.
\end{equation}
The complexity is measured in terms of the following date of the
input:~${n+\ell}$ and the number~$L$ of arithmetic operations needed
to compute the numerators and denominators of~$F$.
\subsection{Differential Algebraic Setting}
\label{sec:DifferentialAlgebraicSetting}
We use differential algebra, founded by \textsc{J.F.\ Ritt}, in order
to introduce forthcoming definitions (see~\cite{Ritt1966} for a
complete description).
\par
The differential algebra~${\mathbb{K}\{t,X,\Theta\}}$ is
the~$\mathbb{K}$-algebra of multivariate polynomials defined by the
infinite set of indeterminates~${\{t,\Theta,X^{(j)} |\, \forall j \in
  \mathbb{N}\}}$ and equipped with time
derivation~${\textup{d}/\textup{d}t}$ denoted by~$\mathcal{L}$. Thus,
for any~$y$ in~${\mathbb{K}\{t,X,\Theta\}}$, relations~${\mathcal{L}
  y^{(i)} = y^{(i+1)}}$ hold.
\par
System~(\ref{eq:ODS}) defines a prime differential ideal~$I$ of the
algebra~${\mathbb{K}\{t,X,\Theta\}}$. This ideal encodes all relations
between coefficients of power series solutions of~$\Sigma$.  In the
sequel, we are going to focus our attention on the quotient
differential fraction field~${\mathbb{K}\{t,X,\Theta\}/I}$ ---denoted
by~$\mathcal{K}$---associated to~$I$.
\paragraph{Derivations vector space.}
Let us recall that derivations acting on~$\mathbb{K}(t,X,\Theta)$ form
a vector space over~$\mathbb{K}(t,X,\Theta)$ denoted
by~${\textup{Der}(\mathbb{K}(t,X,\Theta)/\mathbb{K})}$ and equipped
with a canonical base given by elementary derivations:
\begin{equation}
  \label{eq:DerivationBase}
  \left\lbrace \frac{\partial\hfill}{\partial t},\ 
    \frac{\partial\hfill}{\partial {x_{i}}},\
    \frac{\partial\hfill}{\partial {\theta_{l}}  }\ \Bigg|\
    1\leq i \leq n,\ 1\leq l \leq \ell\right\rbrace\!.
\end{equation}
This vector space equipped with canonical Lie bracket forms a solvable
Lie algebra (see~\cite{ReidWittkopf2000, LisleReidBoulton1995} and
references therein for some algorithmic tools used in study of Lie
algebra in this context). In the sequel, we are going to consider
other such algebras (see
remark~\ref{rem:SolvableLieAlgebraOfSymmetries}).
\paragraph{Canonical field isomorphism.}  
As our input system defines explicitly a vector field, any high order
derivatives could be rewritten to~$0$th order ones using
relations~(\ref{eq:ODS}). Thus, our differential field~$\mathcal{K}$
is isomorphic to the differential field~${\mathbb{K}(t,X,\Theta)}$
equipped with the following formal Lie derivation
in~${\textup{Der}(\mathbb{K}(t,X,\Theta)/\mathbb{K})}$:
\begin{equation}
\label{eq:FirstOrderDerivation}
 \mathcal{D} := \frac{\partial\;}{\partial t} + \sum^{n}_{i=1}
f_{i}\frac{\partial\;\;}{\partial x_{i}}.
\end{equation}
We are going to use directly the canonical isomorphism between
differential fields~${(\mathcal{K},\mathcal{L})}$
and~${(\mathbb{K}(t,X,\Theta),\mathcal{D})}$. All forthcoming
developments are based on derivation~$\mathcal{D}$. Thus, hereafter,
we denote by~$\mathcal{K}$ the field~${\mathbb{K}(t,X,\Theta)}$,
 the set~${(\mathcal{D}f_{1}, \dots, \mathcal{D}f_{n})}$
by~$\mathcal{D}F$ and the composition~${\underbrace{\mathcal{D} \circ
    \dots \circ \mathcal{D}}_{j\ \textup{times}}}$
by~$\mathcal{D}^{j}$.
\paragraph{Formal power series.}
Let us denote by~$\Xi(t,X,\Theta)$ formal power series with
coefficients in~$\mathcal{K}$ that are solutions of~${\dot{\Xi} =
  F(t,\Xi,\Theta)}$ with initial condition~${\Xi(0,X,\Theta):=X}$.
\par
These power series could be define using derivation~$\mathcal{D}$ by
the following formal relations:
\begin{equation}
  \label{eq:PowerSeries}
  \Xi(t,X,\Theta)=\sum_{j \in \mathbb{N}}
  \mathcal{D}^{j} X \; \frac{t^{j}}{j!}.
\end{equation}
\begin{remark}  \label{rem:higherOrdre}
  Higher order derivatives could be considered via  differential
  field~${\mathbb{K}\langle t,X,\Theta\rangle}$ that is the field~${
    \mathbb{K}(t,\Theta,X^{(i)},i\in\mathbb{N})}$ and the formal
  derivation:
    \begin{equation}
      \label{eq:prolongationD}
      \mathcal{D}_{\infty} = 
      \mathcal{D} +\!\! \sum_{j\in \mathbb{N}\setminus\{0,1\}}\sum^{n}_{i=1}
      \mathcal{D}^{j}f_{i}\frac{\partial\hfill }{\partial {x_{i}}^{\!(j)}},
    \end{equation}
    as~$\mathcal{D}_{\infty}X^j \subset\mathcal{K}$, that allows to
    compute higher order derivatives of~$X$.
\end{remark}
In the sequel, we are going to exploit the fact that the differential
field~${(\mathcal{K},\mathcal{D})}$ encodes all formal informations
associated to formal power series solution of~$\Sigma$.
\section{Infinitesimal Generators of Scale Symmetries acting on~$\mathcal{K}$}
\label{sec:InfGenScal}
We are looking for a~$m$-parameters group~$\sigma_{(\lambda_{1},
  \ldots, \lambda_{m})}$ acting on~$\mathcal{K}$ that is a symmetries
group leaving the solution of~$\Sigma$ invariant.  First, we are going
to recall how a derivation
in~${\textup{Der}(\mathcal{K}/\mathbb{K})}$ could be
associated to a scale (resp.\ translation) transformation. Then, we
explicit that such a derivation defines a scale (resp.\ translation)
symmetries of~$\Sigma$.
\subsection{Scale/translation transformation groups and 
  associated derivations vector spaces.}
\label{sec:DerivationsAssociatedToScale}
In order to explicit classical basis of forthcoming computations, let
us consider the following one-parameter group of scale
transformations:
\begin{equation}
  \label{eq:SystemShape}
  \sigma_\lambda:
  \begin{array}{ccr} 
    t & \rightarrow & \lambda^{\alpha_{t}}\, t, \\
    x_{1} & \rightarrow & \lambda^{\alpha_{x_{1}}} \, x_{1},\\
    & \vdots & \\
    x_{n} & \rightarrow & \lambda^{\alpha_{x_{n}}}\, x_{n},
  \end{array} 
  \qquad
  \begin{array}{ccc} 
    \theta_{1} & \rightarrow & \lambda^{\alpha_{\theta_{1}}} \theta_{1}, \\
    & \vdots & \\
    \theta_{\ell} & \rightarrow & \lambda^{\alpha_{\theta_{\ell}}} \theta_{\ell},
  \end{array} 
\end{equation}
the group parameter is denoted by~$\lambda$ and taken in a
field~$\mathbb{F}$; exponents~$\alpha_{i}$ are in a constant field
($\mathbb{Q}$ for example) and they define considered group.
\begin{remark}  \label{rem:translationVerhulst}
  For translation, we consider analogous expressions~${y\rightarrow y
    + \alpha_y \lambda}$ with~$y$ in~$(t,X,\Theta)$. In
  Example~\ref{ex:VerhulstLinearPredation}, we already considered the
  one-parameter translation given by the exponents set defined by
  relations~${ \alpha_a=\alpha_c=1}$
  and~${\alpha_t=\alpha_x=\alpha_b=0}$ that leads to
  transformations~$${ \sigma_\lambda: t \rightarrow t,\; x \rightarrow
    x,\; a \rightarrow a+\lambda, \; b \rightarrow b,\; c \rightarrow
    c+\lambda.}$$
\end{remark}
\par
Actions of these groups could be defined on power series~$\Xi$
solution of~$\Sigma$ (i.e.\ they remain invariant under group
action~${\Xi(\sigma_\lambda(t,X,\Theta))=\Xi(t,X,\Theta)}$) and
on~$\mathbb{K}(t,X,\Theta)$. We consider this last case in the sequel.
\paragraph{Infinitesimal generators vector space.}
When a transformation group is connected---as it is always the case
for transformations considered in this paper---each symmetries of
group~(\ref{eq:SystemShape}) is associated to an infinitesimal
generator (for geometric description, see~\S~2 in~\cite{Olver1993}).
It is a derivation on~$\mathcal{K}$ in the vector space~$\mathbf{S}$
generated in~${\textup{Der}(\mathcal{K}/\mathbb{K})}$ by
the derivation:
\begin{equation}
  \label{eq:ScaleInfinitesimalGenerator} 
  \mathcal{S} = 
  \sum_{y\in (t,X,\Theta)} \alpha_{y} y \frac{\partial\hfill}{\partial y},
\end{equation}
with the~$\alpha_{y}$ are constant exponents taken
from~(\ref{eq:SystemShape}).  In fact, given any
element~$\mathcal{S}_{\rho}$ in~$\mathbf{S}$ equal
to~$\rho\mathcal{S}$ with~$\rho$ an element of a constant field, one
can construct a scale transformation using exponential map as follow:
\begin{equation}
  \label{eq:GroupFromDer}
  \sigma_{\lambda}:
  \begin{array}[t]{ccc}
    \mathcal{K}&\rightarrow&\mathcal{K}(\lambda) \\
    y &\rightarrow& \sum_{i\in\mathbb{N}}{\mathcal{S}_{\rho}}^{\!i}(y)/i!
  \end{array}
\end{equation}
if exponential~${\lambda:=\exp(\rho)}$ is defined. At opposite, one
can determine a derivation associated to the
application~$\sigma_{\lambda}$ but we are not going to use this fact
in the sequel.
\begin{remark}  
  In the case of a~$m$-parameters group, the vector space~$\mathbf{S}$
  is generated by~$m$
  derivations~${\mathcal{S}_{1},\ldots,\mathcal{S}_{m}}$ of the same
  type then~(\ref{eq:ScaleInfinitesimalGenerator}). More
  precisely~$\mathbf{S}$ is a Lie algebra
  s.t.~${[\mathcal{S}_i,\mathcal{S}_j]=0}$.
\end{remark}
\begin{remark}  
  Same considerations are true for translation but in this case
  infinitesimal generators are:
  \begin{equation}
    \label{eq:TranslationInfinitesimalGenerator} 
    \mathcal{T} = 
    \sum_{y\in (t,X,\Theta)} \alpha_{y} \frac{\partial\hfill}{\partial y}.
  \end{equation}
  The vector space of infinitesimal generators associated to a
  translation symmetries group is denoted by~$\mathbf{T}$ in the
  sequel.
\end{remark}
\subsection{Constraint on infinitesimal generators of scale and 
  translation symmetries --- Lie symmetry determining equation}
\label{sec:ConstraintonEulerDerivation}
In order to compute derivations~(\ref{eq:ScaleInfinitesimalGenerator})
and thus, the associated scale symmetries group, we use the classical
property that a derivation~$\mathcal{C}$ associated to a symmetry of a
differential system commutes with the induced derivation~$\mathcal{D}$
(in that case, the derivation~$\mathcal{C}$ is called a symmetry
of~$\mathcal{D}$ by extension).  This fact can be state using Lie
bracket as follows:
\begin{equation}
  \label{eq:commutation}
  [\mathcal{C},\mathcal{D}]:= \mathcal{C}\circ\mathcal{D} - 
  \mathcal{D}\circ\mathcal{C}=\lambda\mathcal{D},
\end{equation}
the parameters~$\lambda$ is a constant; it is different from~$0$ if
considered symmetry does act on times and is~$0$ otherwise.
\paragraph{Infinitesimal conditions defining symmetries groups.}
For reader's convenience, we derive from~(\ref{eq:commutation})
infinitesimal conditions that a derivation is an infinitesimal
generators of a scale (resp.\ translation) symmetry (see \S~2.5
in~\cite{Olver1993} for a presentation of these infinitesimal
conditions based on jet space and prolongation
and~\cite{ChebTerrabRoche1998} for resolution algorithms based on
them).
\begin{lemma} ---
  \label{lem:infinitesimalConditions}
  Infinitesimal conditions of order~$0$ that an infinitesimal
  generator given by~${\sum_{y\in (t,X,\Theta)} \alpha_{y} y
    {\partial}/{\partial y}}$ defines a scale symmetry of ordinary
  differential system~$\Sigma$ are:
  \begin{equation}
    \label{eq:ScaleInfinitesimalCondition}
    \sum_{y\in(t,X,\Theta)} \alpha_y y \frac{\partial f_{i}}{\partial y} +
    (\alpha_t - \alpha_{x_i})f_i =0,\
    \textup{for}\ i=1,\ldots,n.
  \end{equation}
  The infinitesimal conditions of order~$0$ that an infinitesimal
  generator~$\sum_{y\in (t,X,\Theta)} \alpha_{y} {\partial}/{\partial
    y}$ defines a translation symmetry of ordinary differential
  system~$\Sigma$ are:
  \begin{equation}
    \label{eq:TranslationInfinitesimalCondition}
    \sum_{y\in(t,X,\Theta)} \alpha_y  \frac{\partial f_{i}}{\partial y} =0,\
    \textup{for}\ i=1,\ldots,n.
  \end{equation}
\end{lemma}
\begin{proof}
  Recall that we work in vector
  space~$\textup{Der}(\mathcal{K}/\mathbb{K})$, thus one could
  consider each component of relation~(\ref{eq:commutation}) on
  element of canonical base~(\ref{eq:DerivationBase}). Thus,
  coefficient of element~$\partial/\partial t$
  is~${\mathcal{S}\mathcal{D}t - \mathcal{D}\mathcal{S}t = \lambda
    \mathcal{D}t}$ and we deduce that the equality~${\lambda =
    \alpha_t}$ holds. Then, remark that the~$n$ other element's
  coefficient~${[\mathcal{S},\mathcal{D}]x_{i} = \lambda
    \mathcal{D}x_{i}}$ of canonical base could be expanded as:
  \begin{equation}
    \label{eq:ScaleInfinitesimalCondition3}
    \alpha_{x_i}f_i\ -\sum_{y\in(t,X,\Theta)} \alpha_y y \frac{\partial f_{i}}{\partial y}
    = \lambda f_i,\    \textup{for}\ i=1,\ldots,n.
  \end{equation}
  Infinitesimal conditions of order~$0$ are obtained by
  replacing~$\lambda$ by~$\alpha_t$ in these relations.  The proof of
  the second assertion is similar.
\end{proof}
\paragraph{Example~\ref{ex:Murray2002p88} (revisited) ---}  These
infinitesimal conditions associated to example~2 are given by the
matricial relation~${MA=0}$ with matrix~$M$ and vector~$A$ defined in
figure~\ref{fig:Matrix}.
\begin{figure*}[htbp]
  \centering \newlength{\tmplength}
\setlength{\tmplength}{\arraycolsep} \setlength{\arraycolsep}{3pt}
\begin{equation}
  \label{eq:Murray2002p88InfiniteConditions}
\begin{array}{c}
  A := (
  \begin{array}{ccc cccccc}
 \alpha_t &\alpha_n &\alpha_p & \alpha_r & \alpha_{k_1}& \alpha_{k_2}& \alpha_h& \alpha_s& \alpha_e
  \end{array} )
  \\[\medskipamount]
 M:= \left(
  \begin{array}{ccccccccc}
 0& \left({\frac {{k_{2}}p}{\left( n+e \right)^{2}}} 
   - {\frac {r}{{k_{1}}}} \right)rn &
 -{\frac {{k_{2}p}}{n+e}}& r-{\frac {rn}{{k_{1}}}} &{\frac {r{n}}{{{k_{1}}}}}
&-{\frac {pk_{2}}{n+e}}&0&0&{\frac {{k_{2}}pe}{\left( n+e \right)^{2}}} \\
0&{\frac {h{p}}{{n}}}& -{\frac {hp}{n}}&0&0
&0&-{\frac {h{p}}{n}}&  1-{\frac {hp}{n}} &0
  \end{array}
  \right)
\end{array}
\end{equation}
  \caption{Matrix defining infinitesimal conditions associated to example~2}
  \label{fig:Matrix}
 \setlength{\arraycolsep}{\tmplength}
\end{figure*}
\paragraph{Exponents vector space.}  Considered as
relations with coefficients in~$\mathbb{K}(t,X,\Theta)$, the above~$0$th
order infinitesimal conditions are not apparently sufficient to define
completely the vector space~$\mathrm{S}$ (resp.~$\mathrm{T}$) of
exponents associated to derivation vector space~$\mathbf{S}$
(resp.~$\mathbf{T}$) presented in
Section~\ref{sec:DerivationsAssociatedToScale} and thus to determine
searched symmetries groups.
\par
In fact, there is~${(n+\ell+1)}$ unknowns and~$n$ relations; hence
there is apparently no enough relations to define a basis
of~$\mathrm{S}$.
\begin{remark}  \label{rem:VectorSpaceDef}
  Nevertheless, this vector space is well defined because
  derivations~$\mathcal{D}$ and~$\mathcal{S}$ could be prolongated by
  the derivation~$\mathcal{D}_{\infty}$ defined by
  formula~(\ref{eq:prolongationD}) (see remark~\ref{rem:higherOrdre})
  and by~$\mathcal{S}_{\infty}$ defined as follow:
\begin{equation}
  \label{eq:prolongationS}
   \mathcal{S}_{\infty} = \mathcal{S} +
   \sum_{j\in \mathbb{N}^\star}\sum_{y\in X} (\alpha_{y} - j\alpha_{t}) y^{(j)}
   \frac{\partial\hfill}{\partial y^{(j)}}.
\end{equation}
These derivations act on higher order derivatives of initial states
variables i.e.\ on~$\mathbb{K}\langle t,X,\Theta \rangle$; The first part
of~$\mathcal{S}_{\infty}$ is related to relationship between classical
twisted Euler derivations~(\ref{eq:ScaleInfinitesimalGenerator}) and
scaling; the second one is related to the fact that
  \begin{equation}
    \label{eq:GroupActionOnDerivatives}
    \sigma_{\lambda}\!\left(
      \frac{\textup{d}^{j} x_{i}}{{\textup{d} t}^j}
      \right)=     \lambda^{(\alpha_{x_{i}}-j\alpha_{t})} 
      \frac{\textup{d}^j x_{i}}{\textup{d} t^j}, \ j \in \mathbb{N},\ i=1,\ldots,n.
  \end{equation}
  Applying manipulations of lemma~\ref{lem:infinitesimalConditions} on
  prolongated derivations, we obtain enough relations to define
  desired vector space~$\mathrm{S}$ by considering prolongated base
  elements~$\partial^j/\partial {x_i}^{\!j}$. To be more precise, the
  following higher order infinitesimal conditions hold:
  \begin{equation}
    \label{eq:JetLinearSystem}
    \sum_{y\in (t,X,\Theta)}\!\!\! \alpha_{y} y 
    \frac{\partial \mathcal{D}^j x_{i}}{\partial y}
    - (\alpha_{x_{i}} - j \alpha_t) \mathcal{D}^j x_{i} = 0,
  \end{equation}
  for~$i$ in~${(1,\ldots,n)}$ and~$j$ in~$\mathbb{N}$. From this
  infinite set of relations, one can deduce bases of~$\mathrm{S}$ when
  this vector space is not reduced to~$0$ (this is generically the
  case, but as shown in Section~\ref{sec:DimensionalAnalysis}, we
  considered here non generic differential systems coming from
  biology, physics, etc).
\end{remark}
The dimension~$m$ of~$\mathrm{S}$ gives the number of parameters of
our scales symmetries group; once a basis of~$\mathrm{S}$ is chosen,
each of its vector is associated to a one parameter group of scale
symmetries and vector's coefficients allows to determine
exponents---the~$\alpha$s in~(\ref{eq:SystemShape})---of this group.
\par 
One can choose~$m$ of the components of the~$\alpha_i$ quite
arbitrarily. We shall actually want this arbitrariness to bear on the
components corresponding to~$\Theta$ (see
Section~\ref{sec:ComputationOfRationalInvariants}).
\begin{remark}             
  All forthcoming computations are based and devoted to this vector
  space but we are going to see that there is no need to derive
  supplementary infinitesimal conditions from prolongated
  derivations~(\ref{eq:prolongationD}) and~(\ref{eq:prolongationS}).
\end{remark}
\section{Algorithm}
\label{sec:Algorithm}
\subsection{Infinitesimal generators computation}
\label{sec:SeminumericalAlgorithm}
Let us recall that infinitesimal conditions presented in
lemma~\ref{lem:infinitesimalConditions}, show that computation of
considered symmetries groups is associated to computation of a kernel
of a matrix of size~${n\times (n+\ell+1)}$ with coefficients in the
field~$\mathbb{K}(t,X,\Theta)$ (this kernel defines vector
space~$\mathrm{S}$ presented at end of
Section~\ref{sec:ConstraintonEulerDerivation}).
\begin{remark}  
  Fortunately, the vector space~$\mathrm{S}$ is defined over a
  constant field and thus its computation does not require
  computations in~$\mathbb{K}(t,X,\Theta)$. Coordinates~$t,\ X$
  and~$\Theta$ can be specialized to some generic values of a constant
  field in the considered matrix and so, computations could be
  performed numerically with high probability of success (i.e.\ 
  computation failed when specialization are done in a Zariski closed
  set).
\end{remark}
\paragraph{Multiple specializations vs.\ higher
  order computations.}  In order to compute a basis of vector
space~$\mathrm{S}$ as sketched in remark~\ref{rem:VectorSpaceDef}, one
could specialize coordinates~${t, X,\Theta}$ and their higher order
derivations in formula~(\ref{eq:JetLinearSystem}) i.e.\ compute
numerical a single power series solution of linear variational system
derived from~$\Sigma$ (see appendix~\ref{sec:SeriesBasedComutations}
for more details and~\cite{Sedoglavic2001} for another application of
that principle and more details on this computational strategy). This
process could be done several time at different order.
\par
But instead, one could also perform~${\lceil(n+\ell+1)/n \rceil}$
specializations of coordinates~${t, X,\Theta}$ in~$0$th order
infinitesimal conditions presented in
lemma~(\ref{lem:infinitesimalConditions}) and thus, obtain a square
system that allows to compute a base of~$\mathrm{S}$ (see
appendix~\ref{sec:MainValueTrick}, for an example). Thus, we
consider~${\lceil(n+\ell+1)/n \rceil}$ series computations at
order~$0$ in the sequel.
\paragraph{Example~\ref{ex:Murray2002p88} (continued) ---} 
After~$6$ specialization of matrix~$M$ defined in
figure~\ref{fig:Matrix}, one can compute numerically its kernel and
construct the following matrix:
\begin{equation}
\label{eq:Base}
  K:=\left(
  \begin{array}{rrrrrrrrr}
    0& 1& 1& 0& 1& 0& 0& 0& 1 \\
    -1& 0& 0& 1& 0& 1& 0& 1& 0 \\
    0& 0& -1& 0& 0& 1& 1& 0& 0
  \end{array}
  \right)\!, 
\end{equation}
such that~$MK=0$. Rows of this matrix define the vector
space~$\mathrm{S}$ in~$\mathbb{K}^9$ if this vector space is given by
coordinates~${( \alpha_t ,\alpha_n ,\alpha_p , \alpha_r ,
  \alpha_{k_1}, \alpha_{k_2}, \alpha_h, \alpha_s, \alpha_e)}$.  By
  Section~\ref{sec:InfGenScal}, this is sufficient to retrieve the
  searched~$3$ parameters group~(\ref{eq:ScalingMurray2002p88}) of
  scale symmetries. In fact, the last row of matrix~(\ref{eq:Base}) is
  associated to derivation:
\begin{equation}
  \label{eq:DerMurray2002p88}
  \mathcal{S}_\nu = h \frac{\partial}{\partial h} + k_2 \frac{\partial}{\partial k_2} 
  - p \frac{\partial}{\partial p},
\end{equation}
that defines the one-parameter group of scale symmetries:~${ p
  \rightarrow p/\nu,\ h \rightarrow \nu h,\ k_{2} \rightarrow \nu
  k_{2}}$.
\paragraph{Computation of Gradients.}
Infinitesimal conditions are based on gradient computation of
functions~$F$ defining our input differential system; we show in this
section that these computations could be efficiently performed in our
model of complexity which is described below.
\begin{definition}  ---
 \label{def:SLP}
  Let~$\mathcal{A}$ be a finite set of variables. A
  \emph{straight-line program} over~${k[\mathcal{A}]}$ is a finite sequence of
  assignments~${b_{i} \leftarrow b^{\prime} \circ_{i}
    b^{\prime\prime}}$ s.t.~$\circ_{i}$ is in~${\{+,-,\times
    ,{\div}\}}$ and~${\{b^{\prime},b^{\prime\prime}\}}$ is
  in~${\bigcup^{i-1}_{j=1} \{ b_{j} \} \cup \mathcal{A} \cup k}$.  Its
  complexity of evaluation is measured by its length~$L$, which is the
  number of its arithmetic operations.  Hereafter, we use the
  abbreviation {\sc slp} for straight-line program.
\end{definition}
A {\sc slp} representing a rational expression~$f$ is a program that
computes the value of~$f$ from any values of the ground field such
that every division of the program is possible. The following
constructive results taken form \cite{BaurStrassen1983} allows us to
determine a {\sc slp} representing the gradient of~$f$.
\begin{theorem} 
  \label{prop:SLPProp} 
  Let us consider a straight-line program computing the value of a
  rational expression~$f$ in a point of the ground field and let us
  denote by~$L_{f}$ its complexity of evaluation.  One can construct a
  {\sc slp} of length~$5L_{f}$ that computes the value of~$f$ and of
  its gradient~$\textup{grad}(f)$.
\end{theorem}
\paragraph{Complexity result.}
We gather all elements presented above and used in our computation of
symmetries in the following proposition:
\begin{proposition}
  \label{prop:GroupComputation} 
  Let~$\Sigma$ be a differential system as described in
  Section~\ref{sec:MathematicalFramework}.  There exists an algorithm
  that determines a~$m$-parameters group of scale (resp.\ translation)
  symmetries of~$\Sigma$.  The arithmetic complexity of this algorithm
  is bounded by
  \begin{equation}
    \label{eq:ComplexityResult}
      \mathcal{O}\Big((n+\ell+1)\big(L+(n + \ell + 1)(2n +\ell+1)\big)\Big).
  \end{equation}
\end{proposition}
\begin{proof}
  Using theorem~\ref{prop:SLPProp}, the evaluation complexity of
  linear system derived from infinitesimal
  conditions~(\ref{eq:ScaleInfinitesimalCondition}) is bounded by~$
  \mathcal{O}( (L+(n + \ell + 1)n)n)$. This system should be
  evaluated~${\lceil(n+\ell+1)/n \rceil}$ times on some generic
  specializations elements in a constant field in order to obtain a
  square system. Using Gaussian elimination, a base of the resulting
  system kernel could be computed with~$ \mathcal{O}\big ((n + \ell +
  1)^3\big)$ arithmetic operation. Thus, we obtain our
  complexity~(\ref{eq:ComplexityResult}).
\end{proof}
\begin{remark} 
  The algorithm presented in this section is not probabilistic even if
  it is based on specializations that could lead to computation of
  some spurious symmetries. In fact, it could occurs---with a very
  small probability---that coordinates~$t, X, \Theta$ are specialized
  on several points in the orbit of a symmetries group of~$\Sigma$; in
  this case, some spurious symmetries are obtain.
  \par
  Fortunately, a simple evaluation of computed symmetries on our
  original system allows to show if it is a computational artifact or
  not. The complexity associated to these tests is bounded
  by~$\mathcal{O}((n+\ell+1)L)$.
\end{remark}
\begin{remark}  \label{rem:ManyChoice}
  There is an infinite way to choose a basis of~$\mathrm{S}$. But,
  one can use Lenstra, Lenstra and Lov\'asz' basis reduction algorithm
  in order to obtain a reduced basis in the sense that
  exponent~$\alpha$s are smaller then what could be obtained using
  classical Gram-Schmidt orthogonalization.
\end{remark}
\subsection{Computation of some rational invariants and original system rewritting}
\label{sec:ComputationOfRationalInvariants}
Generally speaking, when a differential system admits a group action
as symmetry, we can rewrite it in terms of the invariants of the group
action.  Fels and Olver revised the moving frame the construction for
(differential) invariants~\cite{Olver1993}. The constructed invariants
allow for a trivial rewriting.  We do not recall the general theory
but rather show how it works on the group actions of interest in this
paper.
\par
While computation of \emph{non-specific} rational invariants i.e.\ a
total description of invariant field for any algebraic group action
could be done using reduced Gr\"obner basis computation
(see~\cite{HubertKogan2006}), we do not need such a general tool
because we restrict ourself to scale and translation transformations.
In fact for scale/translation symmetries, we could restrict ourself to
compute~${\ell-m}$ time independent rational invariants in the
multiplicative group in~$\mathbb{K}(\Theta)$ generated by the set:
\begin{equation}
  \label{eq:MultiplicativeGroupOfInvariants}
  \mathbb{M} := \left \lbrace 
    \theta^{\beta}\ \big|\
    (\theta,\beta) \subset \Theta\times\mathbb{Q},\
    \forall \mathcal{S}\in\mathbf{S},\
      \mathcal{S}{\theta} \neq 0
    \right\rbrace\!.
\end{equation}
A set of generators of this multiplicative group is denoted
by~${\pi_1,\ldots,\pi_{\ell-m}}$.  Furthermore, we restrict ourself to
looking for following kind of time dependent rational invariants:
\begin{equation}
  \label{eq:InvariantsType}
  \pi_t = p_t t, \quad \pi_{x_{i}} = p_i x_{i},\ 1\leq i \leq n,
\end{equation}
where the~$p_i$ are in~$\mathbb{M}$. This arbitrary choice simplify
invariants' computation and is well suited with our purpose; we are
trying to use computed group of symmetries to reduce the number of
parameters and not to reduce the number of equations.
\paragraph{Group action.}
Given~$\sigma_{(\lambda_{1}, \ldots, \lambda_{m})}$ a~$m$-parameters
symmetries group, let us consider its action~$\psi$ on
field~$\mathbb{K}(t,X,\Theta)$:
\begin{equation}
  \label{eq:GroupAction}
  \psi:
  \begin{array}{ccccc}
    \mathbb{F}^m&\!\!\!\!\times\!\!\!\!& \mathbb{K}(t,X,\Theta) & \rightarrow & 
    \mathbb{K}(t,X,\Theta), \\
    (\lambda_{1},\ldots,\lambda_{m})&\!\!\!\!\times\!\!\!\!& y  & \rightarrow & 
    \sigma_{(\lambda_{1},
  \ldots, \lambda_{m})}(y).
  \end{array}
\end{equation}
with~$\mathbb{F}$ a field where parameters could be specified. For
scale symmetries, we could wrote~$\psi$ as the substitution given by:
\begin{equation}
  \label{eq:GroupActionSubstitution}
  \forall y\in (t,X,\Theta), \quad \psi(y)= y\, \Pi_{i=1}^{n} {\lambda_{i}}^{\!a_{y,i}}, 
\end{equation}
where exponents are given by the basis~${ \big\lbrace
  (a_{t,i},a_{x_{1},i}, \ldots, a_{x_{1},i},a_{\theta_{1},i},
  \ldots,a_{\theta_{\ell},i}),\ i=1,\ldots,m \big\rbrace}$
  of~$\mathbf{S}$.
\paragraph{Rational invariant computation.}  By classical canonical homomorphism, the
multiplicative set~${\mathbb{F}^m\times\mathbb{M}}$ could be
considered as a~$\mathbb{Z}$ module of dimension~$2\ell$.  Thus, the
subset of relations~(\ref{eq:GroupActionSubstitution}) involving only
parameters could be represented by the following~$\ell \times 2\ell$
matrix:
\begin{equation}
  \label{eq:InvariantComputation}
  \begin{array}{c}
  \begin{array}{cccccccc}
    \lambda_1 & \ldots & \lambda_m  & \theta_1 &\multicolumn{2}{c}\ldots& \theta_\ell
  \end{array}
  \\[\smallskipamount]
 \left(
  \begin{array}{cccccccc}
    a_{\theta_{1},1} & \ldots & a_{\theta_{1},m} & 1 & 0 &\multicolumn{2}{c}{\ldots}  & 0  \\
    a_{\theta_{2},1} & \ldots & a_{\theta_{2},m} & 0 & 1 &  & &   \\
    \vdots & & \vdots  & \vdots && \ddots &&\vdots\\
    a_{\theta_{\ell-1},1}&\ldots&a_{\theta_{\ell-1},m} &  & & & 1 & 0  \\
    a_{\theta_{\ell},1}&\ldots&a_{\theta_{\ell},m} & 0 & \multicolumn{2}{c}{\ldots} & 0 & 1 
  \end{array}
  \right)\!.
  \end{array}
\end{equation}
One can consider the matrix obtained after a permutation
of~(\ref{eq:InvariantComputation}) lines~:
\begin{equation}
  \label{eq:InvariantComputationPermutated}
  \begin{array}{c}
  \begin{array}{cccccccc}
    \lambda_1 & \ldots & \lambda_m  & \hat{\theta}_1 &\multicolumn{2}{c}\ldots& \hat{\theta}_\ell
  \end{array}
  \\[\smallskipamount]
 \left(
  \begin{array}{cccccccc}
    a_{\hat{\theta}_{1},1} & \ldots & a_{\hat{\theta}_{1},m} & 1 & 0 &\multicolumn{2}{c}{\ldots}  & 0  \\
    a_{\hat{\theta}_{2},1} & \ldots & a_{\hat{\theta}_{2},m} & 0 & 1 &  & &   \\
    \vdots & & \vdots  & \vdots && \ddots &&\vdots\\
    a_{\hat{\theta}_{\ell-1},1}&\ldots&a_{\hat{\theta}_{\ell-1},m} &  & & & 1 & 0  \\
    a_{\hat{\theta}_{\ell},1}&\ldots&a_{\hat{\theta}_{\ell},m} & 0 & \multicolumn{2}{c}{\ldots} & 0 & 1 
  \end{array}
  \right)\!.
  \end{array}
\end{equation}
such in order to ensure that the determinant of the
submatrix~$(a_{\hat{\theta}_{i},j})_{i=1,\ldots m}^{j=1,\ldots,m}$ is
not~$0$.
\par
A Gaussian elimination performed on this matrix and terminated
at~${m+1}$ column position leads to the matrix~:
\begin{equation}
    \label{eq:InvariantComputationAfterGauss}
    \left ( 
      \begin{array}{ccccccccccc}
        1     & 0    &\dots &0 &\gamma_{1,\hat{\theta}_1}& \ldots &\gamma_{1,\hat{\theta}_\ell}\\
        
        0     &\ddots&\ddots&\vdots&\vdots& & \vdots\\
        \vdots&\ddots&\ddots&0     &\vdots& & \vdots \\
        0     &\dots & 0    &1     &\gamma_{m,\hat{\theta}_1}& \ldots &\gamma_{m,\hat{\theta}_\ell}\\
        0     &\multicolumn{2}{c}{\ldots} &0     &\beta_{m+1,\hat{\theta}_1}& \ldots & \beta_{m+1,\hat{\theta}_\ell} \\
        \vdots&      &      &\vdots&\vdots& &\vdots\\
        0     &\multicolumn{2}{c}{\ldots}&0     & \beta_{\ell,\hat{\theta}_1} & \ldots& \beta_{\ell,\hat{\theta}_\ell}
      \end{array}
    \right ).
\end{equation}
This computation is sufficient to determine the following generators
of the multiplicative set~(\ref{eq:MultiplicativeGroupOfInvariants})
of rational invariants:
\begin{equation}
  \label{eq:InvariantSet}
  \sigma_{(\lambda_{1}, \ldots, \lambda_{m})}\!\left(\prod_{j=1}^{\ell}
      {\hat{\theta}_j}^{\beta_{m+1,\hat{\theta}_j}}\!\right)=
    \prod_{j=1}^{\ell}{\hat{\theta}_j}^{\beta_{m+1,\hat{\theta}_j}},
    \ldots,
  \sigma_{(\lambda_{1}, \ldots, \lambda_{m})}\!\left(\prod_{j=1}^{\ell}
      {\hat{\theta}_j}^{\beta_{\ell,\hat{\theta}_j}}\!\right)=
    \prod_{j=1}^{\ell}{\hat{\theta}_j}^{\beta_{\ell,\hat{\theta}_j}}.
\end{equation}
Thus, this elimination construct elements~$\pi_i$ of~$\mathcal{K}$
that are invariant under the action~$\psi$; remark that considering
the whole action of~$\psi$ i.e.\ a~${(n+\ell+1)\times 2(n+\ell+1)}$
matrix, the same process allows to construct time dependent invariants
of type~(\ref{eq:InvariantsType}) but we are not going to do so.
\paragraph{Example~\ref{ex:Murray2002p88} (continued) --- }  In
Section~\ref{sec:SeminumericalAlgorithm}, we determine by numerical
computation, a basis of vector space~$\mathrm{S}$ presented in
matrix~(\ref{eq:Base}); taking the~$6$ last columns of this matrix,
one obtains the~$3$ first columns of the following matrix:
\begin{equation}
  \label{eq:Murray2002p88InvariantComputation}
\begin{array}{c}
  \left(
  \begin{array}{ccc cccccc}
    0 & 1 & 0 & 1 & 0 & 0 & 0 & 0 & 0  \\
    1 & 0 & 0 & 0 & 1 & 0 & 0 & 0 & 0  \\
    0 & 1 & 1 & 0 & 0 & 1 & 0 & 0 & 0  \\
    0 & 0 & 1 & 0 & 0 & 0 & 1 & 0 & 0  \\
    0 & 1 & 0 & 0 & 0 & 0 & 0 & 1 & 0  \\
    1 & 0 & 0 & 0 & 0 & 0 & 0 & 0 & 1  
  \end{array}
  \right)\!. \\
  \begin{array}{ccc cccccc}
 \mu &\lambda &\nu & r & k_1& k_2& h& s& e
  \end{array}
\end{array}
\end{equation}
After a Gaussian elimination on the first~$3$ columns of above matrix,
we obtain:
\begin{equation}
  \label{eq:Murray2002p88InvariantComputation2}
  \left(
  \begin{array}{cccrrrrrr}
    0 & 0 & 0 & 1 & 0 & -1 & 1 & 0 & 0  \\
    1 & 0 & 0 & 0 & 1 & 0 & 0 & 0 & 0  \\
    0 & 1 & 0 & 0 & 0 & 1 & -1 & 0 & 0  \\
    0 & 0 & 1 & 0 & 0 & 0 & 1 & 0 & 0  \\
    0 & 0 & 0 & 0 & 0 & -1 & 1 & 1 & 0  \\
    0 & 0 & 0 & 0 & -1 & 0 & 0 & 0 & 1  
  \end{array}
  \right)\!.
\end{equation}
The first line of above matrix defines the rational
invariants~${\invariant{h} = {r h}/{k_{2}}}$ and the last line
defines~${\invariant{e} = e/k_{1}}$, etc.
\paragraph{Algebraic counterpart of Frobenius
  theorem in considered cases.}  First, let us remark that the
action~$\psi$ is surjective. In fact, one can consider the group's
parameters~${\lambda_1=\ldots=\lambda_m=\exp(0)=1}$ that correspond to
identity map. This map is associated to derivation~$0$ in~$\mathbf{S}$
by formula~(\ref{eq:GroupFromDer}).
\par\medskip
The arbitrary choice of parameters made at beginning of
Section~\ref{sec:ComputationOfRationalInvariants} is motivated by the
possibility to rewrite our original dynamic~(\ref{eq:ODS}) in order to
reduce the number of parameters; we want to determine an
expression of the dynamic~$\mathcal{D}$ on an intermediate invariant
field~${\mathcal{F}=\mathcal{K}^{\sigma}}$ 
i.e.~$$\mathbb{K} \hookrightarrow \mathcal{F}= 
      \mathbb{K}(\pi_t,\pi_{x_1},\ldots,\pi_{x_n}, \pi_{1},\ldots,\pi_{\ell-m})
      \hookrightarrow \mathcal{K},$$ 
which transcendence dimension w.r.t.~$\mathbb{K}$ is smaller then the
original one but where the number of variables depending on times does
not change.
\par
In the following proposition, we gives an elementary algebraic proof
of a general results for reader's convenience
(see~\cite{HubertKogan2006} for definition, computation and use of
replacement invariants).
\begin{proposition}
  There exists~${\hat{\lambda}_1,\ldots,\hat{\lambda}_m}$
  in~$\mathbb{M}$ such that the following equality holds~:
  \begin{equation}
    \label{eq:pullback}
    \psi^{-1}\left(\mathcal{K}^{\sigma_{(\lambda_1,\ldots,\lambda_m)}}\right) = 
    {\left(\hat{\lambda}_1,\ldots,\hat{\lambda}_m\right)}\times\mathcal{K}.
  \end{equation}
\end{proposition}
\begin{proof}
  This proposition is a reformulation of
  matrix~(\ref{eq:InvariantComputationAfterGauss})'s structure
  inherited from matrix~(\ref{eq:InvariantComputation}). In fact, the
  submatrix~$(\gamma_{i,\hat{\theta}_j})_{i=1,\ldots,m}^{j=1,\ldots,\ell}$
  keep a track of the performed gaussian elimination i.e.\ the
  following equality holds:
  \begin{equation}
    \label{eq:MatrixProperties1}
\left(
  \begin{array}{ccc}
    a_{\hat{\theta}_{1},1} & \ldots & a_{\hat{\theta}_{1},m} \\
    \vdots & & \vdots  \\
    a_{\hat{\theta}_{m},1} & \ldots & a_{\hat{\theta}_{m},m} \\
  \end{array}
  \right)\!\!\!    
    \left ( 
      \begin{array}{ccc}
        \gamma_{1,\hat{\theta}_1}& \ldots &\gamma_{1,\hat{\theta}_\ell}\\
        \vdots& & \vdots\\
        \gamma_{m,\hat{\theta}_1}& \ldots &\gamma_{m,\hat{\theta}_\ell}
      \end{array}
    \right )\!\! =\!\! 
  \left ( 
    \begin{array}{ccccccccccc}
      1     & 0    &\dots &0 &0 & \ldots &0\\
      0     &\ddots&\ddots&\vdots&\vdots& & \vdots\\
      \vdots&\ddots&\ddots&0     &\vdots& & \vdots \\
      0     &\dots & 0    &1     &0& \ldots &0
    \end{array}
  \right)\!.
  \end{equation}
  If, for~${i=1,\ldots,m}$, we
  define~${\hat{\lambda}_i:=\prod_{j=1}^{\ell}
    {\hat{\theta}_j}^{-\gamma_{i,\hat{\theta}_j}}}$ and notice that
  the above equality shows that the following relations hold:
  \begin{equation}
    \label{eq:MapParamto1}
  \sigma_{(\hat{\lambda}_1,\ldots,\hat{\lambda}_m)}(\hat{\theta}_i)
  = \hat{\theta}_i \prod_{h=1}^{m} 
  \left( \prod_{j=1}^{\ell} 
    {\hat{\theta}_j}^{-\gamma_{h,\hat{\theta}_j}}\right)^{\!a_{\hat{\theta}_i,h}}
  =\hat{\theta}_i\prod_{j=1}^{\ell} 
    {\hat{\theta}_j}^{-\sum_{h=1}^{m} a_{\hat{\theta}_i,h}\gamma_{h,\hat{\theta}_j}}
  = \hat{\theta}_i {\hat{\theta}_i}^{-1}=1.
  \end{equation}
  Therefore, there exists a
  subset~${(\hat{\theta}_{1},\ldots,\hat{\theta}_{m})}$ of parameter's
  set and a subset~${(\hat{\lambda}_1,\ldots,\hat{\lambda}_m)}$
  of~$\mathcal{K}$ such that the
  relations~${\sigma_{(\hat{\lambda}_1,\ldots,\hat{\lambda}_m)}(\hat{\theta}_i)=1}$
  hold for~${i=1,\ldots,m}$. The same type of result is valid for the
  submatrix~$(\beta_{i,j})_{i=m+1,\ldots,\ell}^{j=1,\ldots,\ell}$
  of~(\ref{eq:InvariantComputationAfterGauss}):
\begin{equation}
    \label{eq:MatrixProperties}\scriptstyle
\setlength{\tmplength}{\arraycolsep} \setlength{\arraycolsep}{2pt}
    \left(\!\!
      \begin{array}{ccc}
        a_{\hat{\theta}_{m+1},1} & \ldots & a_{\hat{\theta}_{m+1},m} \\
        \vdots & & \vdots  \\
        a_{\hat{\theta}_{\ell},1} & \ldots & a_{\hat{\theta}_{\ell},m} \\
      \end{array}\!\!
  \right)\!\!    
  \left ( \!\!
    \begin{array}{ccc}
      \gamma_{1,\hat{\theta}_1}& \ldots &\gamma_{1,\hat{\theta}_\ell}\\
      \vdots& & \vdots\\
      \gamma_{m,\hat{\theta}_1}& \ldots &\gamma_{m,\hat{\theta}_\ell}
    \end{array}\!\!
  \right ) =
    \left ( \!\!
    \begin{array}{ccccccccccc}
      \scriptstyle
        -\beta_{m+1,\hat{\theta}_1}& \ldots & 
        \scriptstyle
        1-\beta_{m+1,\hat{\theta}_m+1} & \ldots & 
        \scriptstyle
        -\beta_{m+1,\hat{\theta}_\ell} \\
        \vdots&      &      \vdots&\ddots &\vdots \\
        \scriptstyle
        -\beta_{\ell,\hat{\theta}_1} & \ldots& 
        \scriptstyle
        -\beta_{m+1,\hat{\theta}_m+1} & \ldots & 
        \scriptstyle
        1-\beta_{\ell,\hat{\theta}_\ell}
      \end{array}\!\!
    \right).
\setlength{\arraycolsep}{1pt}
\end{equation}
This matricial relation prove that, for~${i=m+1,\ldots,\ell}$ the
following equalities hold:
\begin{equation}
  \label{eq:map2invariant}
  \sigma_{(\hat{\lambda}_1,\ldots,\hat{\lambda}_m)}(\hat{\theta}_i)
  = \hat{\theta}_i \prod_{h=1}^{m} 
  \left( \prod_{j=1}^{\ell} 
    {\hat{\theta}_j}^{-\gamma_{h,\hat{\theta}_j}}\right)^{\!a_{\hat{\theta}_i,h}}
  =\hat{\theta}_i\prod_{j=1}^{\ell} 
  {\hat{\theta}_j}^{-\sum_{h=1}^{m} a_{\hat{\theta}_i,h}\gamma_{h,\hat{\theta}_j}}
  =\prod_{j=1}^{\ell}{\hat{\theta}_j}^{\beta_{m+1,\hat{\theta}_j}}
 \end{equation}
 To conclude, remark that the same properties hold for time dependent
 variables.
\end{proof}
Thus, after Gaussian elimination performed
on~(\ref{eq:InvariantComputation}), we obtain a description
of~$\mathcal{K}^{\sigma}$ and an
application~$\sigma_{(\hat{\lambda}_1,\ldots,\hat{\lambda}_m)}$ that
maps~$\mathcal{K}$ to~${\mathcal{K}^{\sigma}}$.  In fact, This action
allows to determine rational invariants and rewrite original system in
an set of invariant coordinates with a reduce number of parameters.

%
\paragraph{Example~\ref{ex:Murray2002p88} (continued) --- }
The second, third and fourth lines of
matrix~(\ref{eq:Murray2002p88InvariantComputation2}) give the
specialization~(\ref{eq:SpecializationMurray2002p88}) of parameters
that allows to determine time dependent invariants and system
rewriting.
We summarized computation done in this section by the following
proposition:
\begin{proposition} 
  Computation of rational invariants defined
  in~(\ref{eq:MultiplicativeGroupOfInvariants}) and
  by~(\ref{eq:InvariantsType}) could be performed by Gaussian
  elimination and thus its complexity is bounded
  by~$\mathcal{O}\big(\ell^3\big)$.
\end{proposition}
\begin{remark}  
  The computation of rational invariants presented in this section
  suppose that the considered symmetries groups is composed of
  dilatation. The same consideration holds for translation. In fact,
  the group action is in this case
  \begin{equation}
    \label{eq:GroupActionSubstitutionAdd}
    \forall y\in (t,X,\Theta), \quad \psi(y)=y+\sum_{i=1}^{n}{a_{y,i}}{\lambda_{i}}, 
  \end{equation}
  and is thus linear; that allows exactly the same type of
  computation.
\end{remark}
\begin{remark} \label{rem:SolvableLieAlgebraOfSymmetries}
  If~$\mathcal{S}$ and~$\mathcal{T}$ are two symmetries
  of~$\mathcal{D}$, the Lie bracket~$[\mathcal{S},\mathcal{T}]$ is
  also a symmetry of~$\mathcal{D}$ by Jacobi identity. Thus,
  symmetries form a Lie algebra and if~$\mathcal{S}_i$ is
  in~$\mathbf{S}$ and~$\mathcal{T}_i$ is in~$\mathbf{T}$, we have the
  classical facts that~$[\mathcal{S}_i, \mathcal{S}_j]=0$
  and~$[\mathcal{T}_i, \mathcal{T}_j]$ and~$[\mathcal{S}_i,
  \mathcal{T}_j]$ are in~$\mathbf{T}$ i.e.\ induced derivation is of
  the same type then~(\ref{eq:TranslationInfinitesimalGenerator}).
  Typically, scaling are symmetries of translation's invariant while 
  the opposite is not true.
  \par
  Thus,~$\mathcal{S}$ and~$\mathcal{T}$ generate a solvable Lie
  algebra and above commutation relations show that we have to use
  translation symmetries first in our algorithm to reduce parameter's
  number and then use scale symmetries (see~\S~$2.5$
  in~\cite{Olver1993}).
\end{remark}
\begin{remark} 
  As mentioned in remark~\ref{rem:ManyChoice} there is some freedom
  in choosing the components of a basis of~$\mathrm{S}$
  (resp.~$\mathrm{T}$), the set of exponents that define a scaling
  (resp.\ translation) symmetry of a differential system.  We shall
  always try to have the freedom on the components of~$\alpha$ that
  correspond to~$\Theta$. This is achieved by placing correctly the
  unknown when solving the system by Gaussian elimination; this fact
  allows user to choose parameters to eliminate.  When this is not
  achievable, the general Lie method's could be use to
  solve---partially---the system by quadrature.
\end{remark}
\section{Conclusion and future work}
\label{sec:Conclusion}
In this paper, we consider the computation of scale and/or
transformation group that are symmetries of an ordinary differential
system and the determination of some of their invariants.  We use
these groups and associated invariants in order to rewrite the
ordinary differential system in an set of invariant coordinates with
less parameters.  The complexity of this process is polynomial in
input's size.
\begin{remark}  
  For the sake of simplicity, we do not include control time dependent
  variables~$U$ in our input system~(\ref{eq:ODS}); if such variables
  occur, the ground field is in practice the differential fraction
  field~${\mathbb{K} \langle U \rangle}$. Computations are performed
  after specialization of variables~$U$ on power series with random
  integer coefficients and which are truncated at order~${n+\ell+1}$.
\end{remark}
Same type of result could be proved for more general symmetries that
allow to reduce further the number of significant parameters as shown
below.
\begin{example}  \label{ex:FitzHughNagumo}
  Let us consider a FitzHugh Nagumo model (see~\S~7
  in~\cite{Murray2002}) defined as follow:
  \begin{equation}
    \label{eq:FitzHughNagumoModel}
    \dot{a}=\dot{b}=\dot{c}=\dot{d}=0,\quad
    \textup{d}x/\textup{d}t = c(x-x^3/3 - y + d),\quad
        \textup{d}y/\textup{d}t = (x+ a - b y)/c.
  \end{equation}
  This system does not have scale or translation symmetries that are
  considered in this paper but one can determine that the derivation:
  \begin{equation}
    \label{eq:FitzHughNagumoSymmetry}
    \frac{\partial }{\partial y} +
    b \frac{\partial }{\partial a} +
     \frac{\partial }{\partial d} 
  \end{equation}
  is an infinitesimal generators of the one-parameter symmetries group
  (which is not of type~(\ref{eq:TranslationInfinitesimalGenerator})):
  \begin{equation}
    \label{eq:FitzHughNagumoTrans}
    y \rightarrow y + \lambda,\quad
    a \rightarrow a + b\lambda,\quad
    d \rightarrow d + \lambda.
  \end{equation}
  Up to our knowledge, there is likely no polynomial time algorithm
  that compute infinitesimal
  generators~(\ref{eq:FitzHughNagumoSymmetry}). In fact, this type of
  symmetries can be found by supposing that all coefficients of seeked
  infinitesimal generators of symetries are rational function of
  parameters; in that case the solution of determining system of
  PDE~(\ref{eq:commutation}) is reduced to the computation of a
  polynomial matrix kernel.  But as done in this paper, one can use
  the following invariant coordinates~${\invariant{y}=y-d}$
  and~${\invariant{a}=a+bd}$ to rewrite
  system~(\ref{eq:FitzHughNagumoModel}) as follow:
  \begin{equation}
    \label{eq:FitzHughNagumoModel2}
    \textup{d}x/\textup{d}t = c(x-x^3/3 - \invariant{y}),\quad
        \textup{d}\invariant{y}/\textup{d}t = (x+ \invariant{a} - b \invariant{y})/c
  \end{equation}
\end{example}
\paragraph{Acknowledgments.}  
The second author is grateful to F.\ Lemaire and M.\ Safey El Din
for many useful suggestions that helped to considerably improve
correctness and presentation of this paper.
\bibliographystyle{acm} \bibliography{HubertSedoglavic2006}
\appendix
\section{Two computation methods}
Considering a classical example taken from biology, we are going to explicit
some computation evoked in Section~\ref{sec:SeminumericalAlgorithm}.
First, we determine a symmetries group using power series approach and
then, we retrieve the same result using just specialization of~$0$th
order infinitesimal conditions.
\subsection{Series based computations}
\label{sec:SeriesBasedComutations}
Let us consider the linear variational system:
\begin{equation}
  \label{eq:LinearVariationalSystem}
  \nabla\quad\left\lbrace
   \begin{array}{ccl}  
\dot{\Xi}&=& F(t,\Xi,\Theta), \\[\smallskipamount]
   \displaystyle
    \frac{\textup{d}\hfill}{\textup{d}t}
    \frac{\partial \Xi}{\partial X} &     \displaystyle = &     \displaystyle

     \frac{\partial F}{\partial X}(t,\Xi,\Theta) \frac{\partial \Xi}{\partial X},
      \\[\medskipamount]     \displaystyle
\frac{\textup{d}\hfill}{\textup{d}t}
     \frac{\partial \Xi}{\partial \Theta} &     \displaystyle = &     \displaystyle
     \frac{\partial F}{\partial X}(t,\Xi,\Theta) \frac{\partial \Xi}{\partial \Theta}+
     \frac{\partial F}{\partial \Theta}(t,\Xi,\Theta),
\end{array} \right.
\end{equation}
with the initial conditions~${{\partial \Xi}/{\partial X} =
  \textup{Id}_{n\times n}}$ and~${{\partial \Xi}/{\partial \Theta} =
  0}$ when~${t=0}$. Power series solutions of this system
are~(\ref{eq:PowerSeries}) and:
\begin{equation}
  \label{eq:SolutionLinearVariationalSystem}
     \frac{\partial\Xi}{\partial X}       = 
     \sum_{j \in \mathbb{N}}
     \frac{\partial \mathcal{D}^{j} X}{\partial X} \; \frac{t^{j}}{j!}, \qquad
     \frac{\partial \Xi}{\partial \Theta} = 
     \sum_{j \in \mathbb{N}}
     \frac{\partial \mathcal{D}^{j} X}{\partial \Theta} \; \frac{t^{j}}{j!}.
\end{equation}
Coefficients of these series are used in generalized infinitesimal
conditions~(\ref{eq:JetLinearSystem}) that allows symmetries
computations.
\par
Hence, one can construct a system of ordinary differential equations
that allows to compute directly a specialization of~${\mathcal{D}^j
  x_{i}}$ and~${\partial \mathcal{D}^j x_{i}/\partial y}$ with~$y$
in~${(X,\Theta)}$.  In fact, on can compute power series solutions of
this system and, the wanted quantities are coefficients of these power
series.
\begin{example}  \label{ex:MichaelisMenten}
  We perform above computations on the following Michaelis Menten's
  equation (see~\S~6.3 in~\cite{Murray2002}):
\begin{equation}
  \label{eq:MichaelisMenten}
  \dot{\xi}=\frac{k_{1} \xi}{k_{2}+\xi} = f(k_{1},k_{2},\xi).
\end{equation}
As~${\partial f/\partial t \equiv 0}$, the associated linear
variational system is:
\begin{equation}
  \label{eq:MMLinearVariationnalSystem}
  \left\lbrace
  \begin{array}{ccc}
    \frac{\textup{d}\hfill}{\textup{d}t}
    \frac{\partial \xi}{\partial x} &=& \frac{k_{1}k_{2}}{(k_{2}+\xi)^2}
    \frac{\partial \xi}{\partial x},\quad \dot{k}_1 = \dot{k}_2 = 0, 
    \\[\medskipamount]
    \frac{\textup{d}\hfill}{\textup{d}t}
    \frac{\partial \xi}{\partial k_{1}} &=& \frac{k_{1}k_{2}}{(k_{2}+\xi)^2}
    \frac{\partial \xi}{\partial k_{1}} + 
    \frac{\xi}{(k_{2}+\xi)},\\[\medskipamount] 
    \frac{\textup{d}\hfill}{\textup{d}t}
    \frac{\partial \xi}{\partial k_{2}} &=& \frac{k_{1}k_{2}}{(k_{2}+\xi)^2}
    \frac{\partial \xi}{\partial k_{2}} -   \frac{k_{1}\xi}{(k_{2}+\xi)^2}.
  \end{array}
  \right.
\end{equation}
Using specializations defined by~${k_{1}=7, k_{2}=2}$
and~${\xi(0)=3}$, at order~$5$ power series
solutions~$\bar{\Xi}(t,3,7,2)$ of above equations are:
\setlength{\tmplength}{\arraycolsep} \setlength{\arraycolsep}{2pt}
\begin{equation}
  \label{eq:MMPowerSeriesSolutions}
  \begin{array}{ccl}
    \xi &=& 3 +\frac{21}{5} t +\frac {147}{125} t^{2}
    -\frac {1372}{3125} t^{3}+\frac {2401}{31250} t^{4} +\mathcal{O}( t^5), 
    \\[\smallskipamount]
    \frac{\partial \xi}{\partial k_{1}} &=& \frac {3}{5} t +\frac{42}{125}  t^{2}
    -{\frac {588}{3125}}{t}^{3}+\frac {686}{15625}{t}^{4}+ \mathcal{O}( {t}^{5}), 
    \\[\smallskipamount]
    \frac{\partial \xi}{\partial k_{2}} &=& -\frac {21}{25}t - \frac {147}{1250}{t}^{2} 
  + {\frac {1029}{3125}}{t}^{3}-{\frac {69629}{312500}}{t}^{4}+\mathcal{O} ( {t}^{5} ), 
    \\[\smallskipamount]
  \frac{\partial \xi}{\partial x} &=& 1+ \frac{14}{25}t - \frac {196}{625} {t}^{2}
  + \frac {686}{9375} {t}^{3}+\frac {16807}{234375} {t}^{4}+ \mathcal{O} ( {t}^{5}).    
  \end{array}
\end{equation}
\setlength{\arraycolsep}{1pt } Using these values, we construct linear
system associated to commutation condition~(\ref{eq:commutation}):\small
  \begin{equation}
    \label{eq:MMLinearSystem}
    \!\left(
    \begin {array}{cccc} 
      -1&0&0&0\\
      \noalign{\medskip}0&-{\frac {63}{25}}&{\frac {21}{5}}&{-\frac {42}{25}}
      \\\noalign{\medskip}{\frac {294}{125}}&-{\frac {2646}{625}}&{\frac {588}{125}}&-{\frac {294}{625}}\\\noalign{\medskip}-{
        \frac {8232}{3125}}&{\frac {4116}{3125}}&-{\frac {16464}{3125}}&{\frac {12348}{3125}}
    \end {array} 
    \right)\!\!
      \left(
   \begin{array}{c}
     \alpha_{t} \\
     \alpha_{x} \\
     \alpha_{k_{1}} \\
     \alpha_{k_{2}}
   \end{array}
   \right)\!\!=\!\lambda\!
      \left(
   \begin{array}{c} 
     1 \\[\smallskipamount]
     {\frac {21}{5}} \\[\smallskipamount]
     {\frac {294}{125}} \\[\smallskipamount]
     -{\frac {8232}{3125}}
   \end{array}
   \right)
  \end{equation}
  \setlength{\arraycolsep}{\tmplength}%
  \normalsize%
  and by a Gaussian elimination, we determine the basis
  of~$\mathrm{S}$ defined by the relations~${\alpha_t =
    -\lambda, \alpha_{k_{1}} = \lambda + \mu}$ and~${\alpha_{k_{2}} = \mu,
    \alpha_x = \mu}$.  Thus, the dimension of vector
  space~$\mathbf{S}$ is~$2$ and the following derivations:
  \begin{equation}
    \label{eq:MMInfGenSol}
    \mathcal{S}_{\lambda} = t \frac{\partial}{\partial t} - k_{1}\frac{\partial}{\partial k_{1}}
    , \quad
    \mathcal{S}_{\mu} = x\frac{\partial}{\partial x} + k_{1}\frac{\partial}{\partial k_{1}}
    + k_{2} \frac{\partial}{\partial k_{2}}
  \end{equation}
  form one of its bases (for the sake of simplicity, higher order
  derivatives are omitted in these derivations).  We deduce
  from~(\ref{eq:GroupFromDer}) that the following~$2$ parameters
  groups:
  \begin{equation}
    \label{eq:MMGroup}
    \begin{array}{ccr} 
      t & \rightarrow & \lambda\, t, \\
      x & \rightarrow & \mu\, x,\\
    \end{array} 
    \qquad
    \begin{array}{ccc} 
      k_1 & \rightarrow & \mu k_1/\lambda, \\
      k_{2} & \rightarrow & \mu k_{2}
    \end{array} 
  \end{equation}
  acts on~(\ref{eq:MichaelisMenten}) and that~${\invariant{t}=k_1 t/k_2}$
  and~${\invariant{x}=x/k_{2}}$ are a convenient set of new coordinates in which
  system~(\ref{eq:MichaelisMenten}) could be rewritten as~${
    \textup{d}\invariant{x}/\textup{d}\invariant{t}=\invariant{x}/(1+\invariant{x})}$.
\end{example}
\subsection{Multiple specialization}
\label{sec:MainValueTrick}
\textit{Example~\ref{ex:MichaelisMenten} (continued) --- } For this
example,~$0$th order infinitesimal
condition~(\ref{eq:ScaleInfinitesimalCondition}) leads to consider the
following vector:
\begin{equation}
  \label{eq:MM1OrderInfCond}
  \left({\frac {{k_{1}}\,x}{{k_{2}}+x}},
    -{\frac {{k_{1}}\,{x}^{2}}{\left( {k_{2}}+x \right)^{2}}},
    {\frac {{k_{1}}\,x}{{k_{2}}+x}},-{\frac {x{k_{1}}\,{k_{2}}}{\left( {k_{2}}+x \right)^{2}}}
  \right),
\end{equation}
Using specialization:
\[
\begin{array}{cc}
x=-2,{ k_1}=10,{ k_2}=-2, &
x=-4,{ k_1}=-7,{ k_2}=1,\\
x=2,{ k_1}=8,{ k_2}=-1, &
x=4,{ k_1}=-2,{ k_2}=1
\end{array}
\]
one can obtain the following matrix associated to infinitesimal
condition~(\ref{eq:ScaleInfinitesimalCondition}): 
\begin{equation}
  \label{eq:MMspecializations}
\left( \begin {array}{cccc} 
  5&-5/2&5&-5/2\\
  -{ {28}/{3}}&{ {112}/{9}}&-{ {28}/{3}}&-{ {28}/{9}} \\
  16&-32&16&16\\
-8/5&{ {32}/{25}}&-8/5&{ {8}/{25}}\end {array} \right)
\end{equation}
whose kernel is defined by vectors~$(1, 1, 0, 1)$ and~$(-1, 0, 1, 0)$
that generates the vector space~$\mathrm{S}$ already determined above
(applying LLL reduction on these vectors, we retrieve exactly the
previous basis).
\par
Computation of numerical power series solutions
of~(\ref{eq:MichaelisMenten})
and~(\ref{eq:MMLinearVariationnalSystem}) performed above could be
done using specialization~(\ref{eq:MMspecializations}). Instead of
considered series solutions associated to a single specialization, one
can consider two or more such series. The series' order needed by our
computation decreases with the number of used specialization.
\par
So, there is no need to compute power series as described above even
if theoretical structures are clearly defined using this approach (see
remark~\ref{rem:VectorSpaceDef}).
\end{document}